\documentclass{svproc}

\usepackage{hyperref}

\usepackage{color}
\usepackage{epsfig}
\usepackage{amssymb}
\usepackage{amsbsy}

\renewcommand{\!}{\negthinspace}
\newcommand{\LA}{\left\langle}
\newcommand{\RA}{\right\rangle}

\def\al{\alpha}

\def\om{\omega}

\def\bz{\bar{z}}

\newcommand{\ta}{{ {a^2}}}
\newcommand{\taa}{{ {a^4}}}

\newcommand{\overset}[2]{\binrel@{#2}%
  \binrel@@{\mathop{\kern\z@#2}\limits^{#1}}}

\newcommand{\tr}[1]{\,{\rm tr}\,#1}

\def\e{{\,\rm e}}
\def\be{\begin{equation}}
\def\ee{\end{equation}}
\def\beq{\begin{equation}}
\def\eeq{\end{equation}}
\def\bea{\begin{eqnarray}}
\def\eea{\end{eqnarray}}
\def\LA{\left\langle}
\def\RA{\right\rangle}
\newcommand{\rf}[1]{(\ref{#1})}
\newcommand{\eq}[1]{Eq.~(\ref{#1})}

\def\hg{{\hat g}}

\def\d{{\rm d}}

\def\i{{\rm i}}

\def\ta{a^2}

\def\om{\omega}

\newcommand{\eps}{{ {\varepsilon}}}
\newcommand{\non}{\nonumber \\*}

\newcommand{\vp}{\varphi}
\newcommand{\p}{\partial}
\newcommand{\bp}{\bar\partial}

\renewcommand{\d}{{\rm d}}
\renewcommand{\i}{{\rm i}}
\newcommand{\blambda}{\bar\lambda}

\renewcommand{\tr}{\mathrm{tr}}

\newcommand{\tit}{\subsection*}

\usepackage{url}

\begin{document}
\mainmatter              
\title{``Massive'' Conformal Field Theories  and
\\ application to Nambu-Goto string}
\titlerunning{Massive Conformal Field Theories and  Nambu-Goto string}  
%
\author{Yuri Makeenko}
\authorrunning{Yuri Makeenko} 
%
\tocauthor{}
\institute{NRC ``Kurchatov Institute''\/-- ITEP, Moscow \\[2mm]
\email{makeenko@itep.ru, yuri.makeenko@gmail.com}}

\maketitle              

\begin{abstract}
I describe the recent progress in understanding diffeomorphism invariant two-dimensional theories 
with dimensionful parameters which have nevertheless traceless improved energy-momentum tensor and turn out to be conformal invariant. An example is the Liouville theory
with the additional four-derivative terms arising from the Nambu-Goto string and telling apart the Nambu-Goto and Polyakov strings. 
Its exact solution in embedding dimension d=4 is consistent in contrast to
the Polyakov string and described by the (4,3) unitary minimal model.\\[2mm]
Talk at Irina Aref'eva Conference
``Perspective Directions in Quantum Field Theory and Superstrings'',
Moscow 12-17 October 2026.
\end{abstract}


\section*{Brief Introduction}

This talk is addressed to the problem of $d=1$ barrier for the existence of strings in
$d$-dimensional embedding space as first found by
Knizhnik-Polyakov-Zamolod\-chi\-kov~\cite{KPZ88} (KPZ) and 
then reproduced by David~\cite{Dav88}, 
Distler-Kawai~\cite{DK89} (DDK) using the Liouville theory.
I shall review this in detail below. 

I describe a (successful) proposal to resolve the problem by adding 
four-derivative terms to the Liouville action. This does not affect the case of the Polyakov
string but for the Nambu-Goto string the four-derivative terms revive and shift the
KPZ-barrier to the value $15-4\sqrt{6}\approx 5.2 >4$. 
This alludes for the Nambu-Goto string to be healthy in $d=4$ and reduced to the (4,3) unitary minimal model.

\section{Four-derivative Liouville theory
 as MCFT}

\tit{Four-derivative Liouville action}

I begin with introducing
{the }{Liouville action} with {}{four-derivative} terms added 
\be
{\cal S}[\vp]=\frac {1}{16 \pi b_0^2} \int \sqrt{\hg} \left[\hg^{ab} \p_a \vp \p_b \vp 
+\eps  \e^{-\vp}\hat\Delta \vp \left(\hat\Delta\vp- G \hg^{ab}\,\partial_a \vp\partial_b \vp \right) +\mu\e^\vp\right] 
\label{4dLiou}
\ee
where
$\hg_{ab}$ --  a (flat) background metric tensor and $\hat \Delta$
 -- the Laplacian for $\hg_{ab}$. 

The action~\rf{4dLiou} is the most general four-derivative one. 
All other possible terms with four derivatives can be reduced to these two modulo
total derivatives (leading to  boundary terms) like
\be
\int \e^{-\vp} \left[(\p\bp \vp)^2-G \p\vp\bp\vp\p\bp\vp\right]=
\int\e^{-\vp} \left[(1+G)(\p\bp \vp)^2-G \nabla\p\vp\bar\nabla\bp\vp\right]
\label{modulo}
\ee
which illustrates stability of the action~\rf{4dLiou} for  $-1\leq G\leq 0$.

The 
 1st  four-derivative term in \eq{4dLiou} is inherited from the Polyakov string
where $G=0$, while the 2nd one with $G=-1/3$ areses from the Nambu-Goto string.
The derivation is given in Sect.~\ref{s:8} below, fixing
$$
  b^2_0=\frac 6{26-d}, \qquad
\eps\sim \Lambda^{-2} 
$$
with $\Lambda$ being a target-space UV cutoff of the string. Notice, this $\eps$ is
dimensionful.

{Classically} the higher-derivative terms vanish for smooth $\vp$ as $\eps\to0$.
Quantumly the quartic derivative in the 1st one
provides a UV cutoff but also an interaction with the coupling $\eps$ 
$\Longrightarrow$ uncertainties $\eps \times \eps^{-1}$ which revive 
${}{\Longrightarrow}$ {}{anomalies}. As I describe  below, the fate of such terms
coming from the 1st and 2nd four-derivative
terms will be different: the former contaminate intermediate
computations but cancel each other in the final results, while the later give
a contribution to observables.


A separate comment is required about the parameter $\mu$ in the last term in \rf{4dLiou}
which is present already for the usual Liouville action. 
 For the Polyakov string ($\Rightarrow$ the usual Liouville action) one may have $\mu\neq0$
 as inherited from the {classical ground state}, while
$\mu=0$ for the Nambu-Goto string as prescribed~\cite{AM16b} in $d>2$ by 
the quantum mean-field ground state.

\tit{Covariant four-derivative action}

The four-derivative {Liouville action}~\rf{4dLiou} is obtaned from {nonlocal} covariant one
\be
{\cal S}[g]=\frac {1}{16 \pi b_0^2} \int \sqrt{g} \left[-R\frac 1\Delta R
+\eps \left(R^2+ G g^{ab}\,\partial_a \frac 1\Delta R\,\partial_b \frac 1\Delta R \right) +\mu \right] 
\label{covar}
\ee
fixing  the {conformal gauge}
\be
g_{ab}=\hg_{ab}\e^\vp,  \qquad \sqrt g R=\sqrt \hg \big(\hat R -\hat \Delta \vp \big)
\label{confog}
\ee
with the scalar curvature $R$ and the background curvature $\hat R$.
Additional terms like
\be
\delta{\cal S}[\vp]=\frac {1}{8 \pi b_0^2} \int \sqrt{\hg} \hat R\vp+\ldots
\label{6}
\ee
also emerge for curved backgrounds.
Notice the double nonlocality of the term with $G$ which will be crucial for what follows.

It becomes clear from \rf{covar} that {anomalies arise as smallness of $\eps $ 
is compensated by a global dilatation 
 (shift of $\vp$)} thanks to {diffeomorphism invariance} of ${\cal S}[g]$.

\tit{Improved Energy-Momentum Tensor}

The symmetric {}{minimal} {}{energy-momentum tensor (EMT)} 
is associated with {minimal coupling} to background gravity 
and reads~\cite{Mak22} ($T_{ab}=-4\pi \delta {\cal S}/\delta \hg^{ab}$)
\bea
&&-4 b_0^2 T_{ab}^{({\rm min})}=\p_a \vp \p_b \vp -\frac 12 g_{ab} \p^c \vp \p_c \vp
 -  {\mu} g_{ab}-\eps \p_a \vp \p_b \Delta \vp -
\eps \p_a \Delta \vp  \p_b \vp  \non&& \hspace*{12mm}
+\eps g_{ab} \partial^c \vp\p_c \Delta \vp+\frac \eps2 g_{ab} (\Delta \vp)^2 
-G \eps \p_a \vp\p_b\vp \Delta \vp 
 + G\frac\eps2\p_a\vp \p_b(\p^c \vp \p_c \vp) \non &&\hspace*{12mm}
+ G\frac\eps2 \p_a (\p^c \vp \p_c \vp)\p_b\vp -G\frac\eps2
g_{ab} \p^c\vp \p_c (\p^d \vp \p_d \vp)
\label{Tabmin}
 \eea
(same as Belinfante's EMT). It is conserved 
but not traceless.

A {nonminimal interaction} with background gravity coming from \eq{confog} 
and shown in \eq{6} is accounted for
by {``Improved'' EMT} as  pointed out by Callan-Coleman-Jackiw~\cite{CCJ70}
and adapted for two dimensions   
in Ref.~\cite{DJ95}. From the action~\rf{covar} we obtain 
 \bea
-4 b_0^2 T_{ab} &=&-4 b_0^2 T_{ab}^{({\rm min})}
-2 (\p_a\p_b-g_{ab}\p^c\p_c )(\vp -\eps \Delta \vp+G\frac\eps2 g^{ab}\p_a \vp \p_b \vp  ) 
\non &&
 + {}{2 G \eps (\p_a \p_b-g_{ab} \p^c\p_c) 
 \frac 1{\Delta} \partial^d(\p_d \vp \Delta\vp) } \hspace*{17mm} \fbox{nonlocal!}
 \label{Tab}
\eea
which is {}{conserved} and {}{traceless} (!) thanks to the classical equation of motion.
Notice the emergence of the nonlocal term in \rf{Tab}.

The tracelessness of Improved EMT implies conformal invariance of the theory.
 It  is the one suitable  for use in ``Massive'' Conformal Field Theory  (MCFT).

\tit{Covariant Pauli-Villars regularization}

{}{Improved} EMT~\rf{Tab} is traceless in spite of $\eps$ and $\mu$ are dimensionful. 
We can analogously consider massive {}{Pauli-Villars} regulators 
\be
{\cal S}^{({\rm reg})} =\frac 1{16\pi b_0^2}\int \sqrt{g}\left[ g^{ab}\partial_a  Y \partial_b Y +M^2 Y^2+
\eps (\Delta Y)^2 -G\eps g^{ab}\partial_a  Y \partial_b Y \Delta \vp \right] 
\label{Sreg}
\ee
which also contribute to Improved EMT
\bea
-4 b_0^2 T_{ab}^{({\rm reg})}&=&\p_a Y \p_b Y -\frac 12 g_{ab} \p^c Y \p_c Y
 -\frac{M^2}2 g_{ab} Y^2-\eps \p_a Y \p_b \Delta Y   -
\eps \p_a \Delta Y  \p_b Y \non
&& \hspace{-2cm} +\eps g_{ab} \partial^c Y\p_c \Delta Y
+\frac \eps2 g_{ab} (\Delta Y)^2
-G \eps \p_a Y\p_b Y \Delta \vp 
 + G\frac\eps2\p_a\vp \p_b(\p^c Y \p_c Y) \non
 && 
 \hspace{-2cm}+ G\frac\eps2 \p_a (\p^c Y \p_c Y)\p_b\vp- G\frac\eps2
g_{ab} \p^c\vp \p_c (\p^d Y \p_d Y) 
 -G\eps (\p_a\p_b-g_{ab} \p^c \p_c) (\p^c Y\p_c Y)
\nonumber   
 \eea
Nevertheless total {}{improved} EMT $T_{ab}+T_{ab}^{({\rm reg})}$
is {}{traceless} thanks to the
classical equation of motion for $\vp$. This is a general property because
\be
T_a^a \propto\hat g^{ab} \frac{\delta {\cal S}[g]}{\delta \hat g^{ab} }= -
\frac{\delta {\cal S}[g]}{\delta \vp }\qquad
\hbox{for}
\qquad
g_{ab}=\hat g_{ab} \e^\vp
\label{Taa}
\ee
Thus {Pauli-Villars'} regulators {}{do not break} conformal symmetry 
 despite they are {}{massive}
({in contrast to QFT without diffeomorphism invariance}).
This explains why Logs do not appear in the calculation of observables below.
 

\section{CFT \'a la KPZ-DDK\label{s:2}}

 \tit{Extension of KPZ-DDK to four-derivative action} 


The conformal weight of $\e^{\vp}$ and the central charge are determined by the
 {operator products} $T_{zz}(z) \e^{\vp(0)}$ and  $T_{zz}(z) T_{zz}(0)$ 
 depicted at the one-loop order as
$$
\includegraphics[width=10cm]{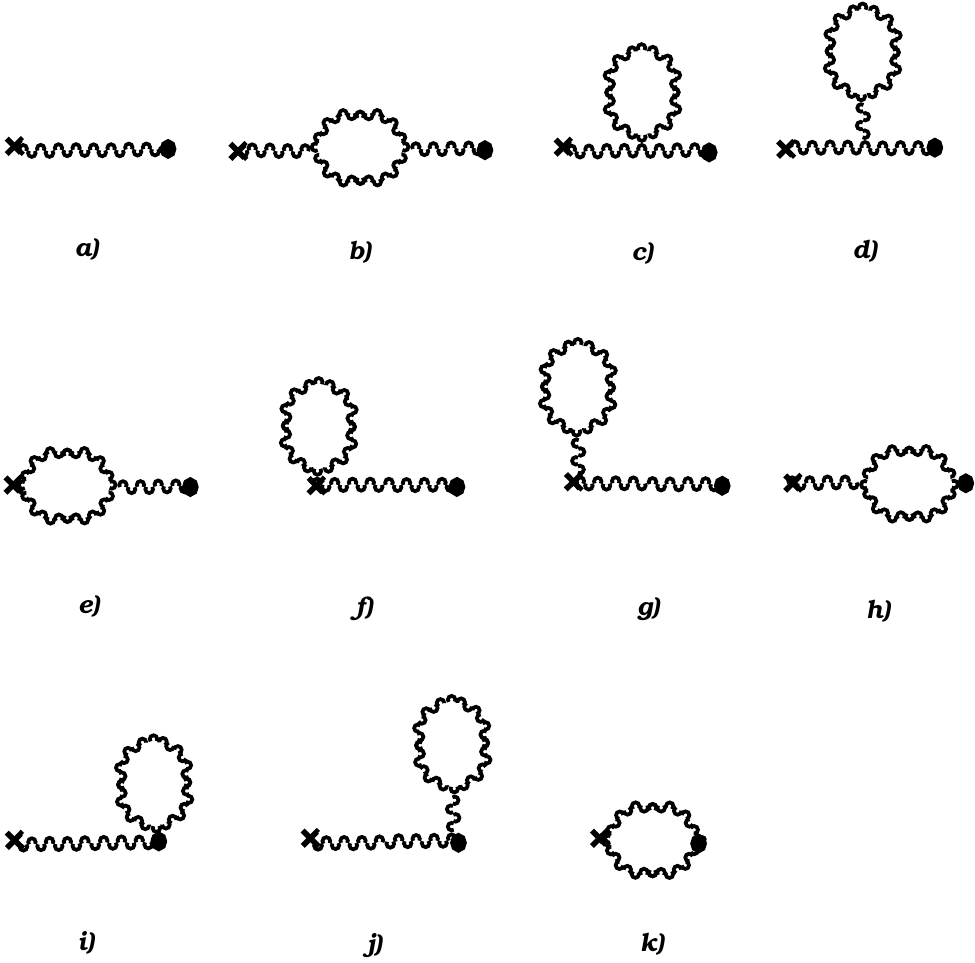} 
$$
The diagrams $b)$ to $d)$ are simply the one-loop correction to the propagaror which results 
in the {renormalization} $b_0^2\to b^2 =b_0^2 + {\cal O}(b_0^4) $. Analogously
the diagrams $e)$ to $g)$ renormalize the linear part of EMT
$$
T^{(\varphi)}_{zz}=
- \frac1{4 b^2}
\left(\partial_z \varphi \partial_z \varphi -2 {q }\partial_z^2 \varphi \right) 
+\ldots ,
 \quad q_0=1\to  {q }=1 +{\cal O}(b_0^2)
$$
and the diagrams $h)$ to $j)$ renormalize
$$
\e^\vp \to \e^{\alpha \vp}, \qquad  \alpha_0=1 \to \alpha =1 + {\cal O}(b_0^2)
$$
or renormalize $T_{zz}(0)$. Thus 
only the diagrams $a)$ and $k)$ actually are to be calculated. 

\tit{Intelligent one-loop approximation} 

We assume that the renormalization preserves the form of the four-derivative action changing
only parameters as just discussed at one loop:
$$
b_0^2  
\to b^2
$$
$$
\sqrt{g} R =\sqrt{\hat g}\,( {q}\hat R -\hat \Delta \vp),\quad q_0=1\to q
$$
$$
g_{ab}=\hg_{ab} \e^{\alpha \vp},\quad \alpha_0=1\to \alpha\equiv 1 
\quad \hbox{(normalization)}
$$
It is motivated by KPZ-DDK and I call it as Intelligent one-loop approximation.
Alternatively, we can fix the normalization choosing $q=1$ like in Ref.~\cite{KN93}.

In the limit $\eps\to0$ only the cross-terms between the nonlocal term in~\rf{Tab} and 1st and 4th
terms in \rf{Tabmin} are nonzero so the background independence yields: 
$$ 
\begin{array}{rl} 
\hbox{ {conformal weight} }&~~~
d^{(\e^{\varphi}) }=  {q } -b^2 =1 \\[2mm] 
\hbox{total  {central charge} }&~~~
{d-26}+ 6 \displaystyle{\frac { {q }^2}{b^2} }+1+6Gq=0 
\end{array}
$$
The emergence of $\delta c^{(\vp)}= 6 {} G q$ in the central charge of $\vp$
is nontrivial and due to the revival of the nonlocal term, where $\Delta^{-1}$ plays
the role of a massless propagator. It thus bypasses the assertion formulated in footnote
on p.~343 of Ref.~\cite{BPZ}:
 \hspace*{3mm}\fbox{\includegraphics[width=11.4cm]{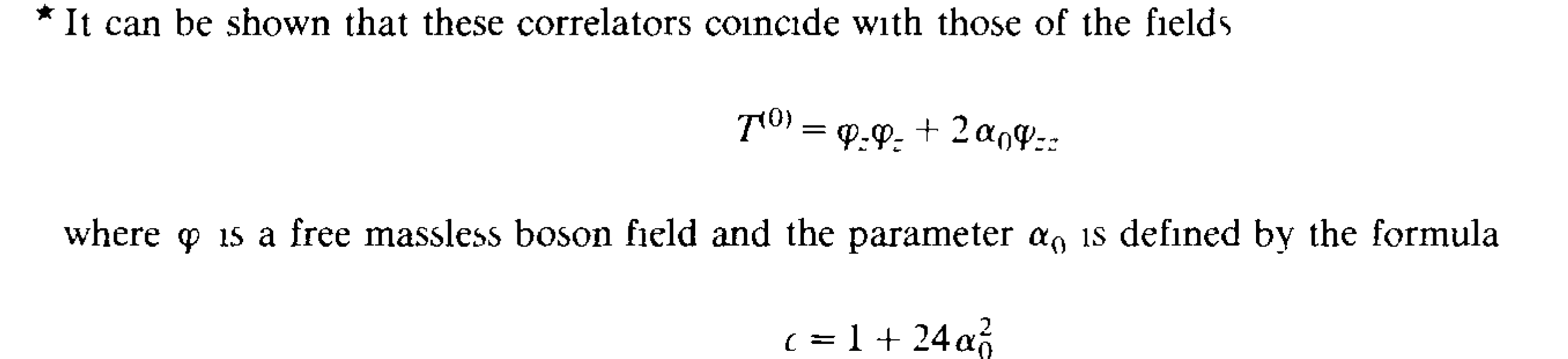}} 

\vspace*{1mm}
{Intelligent one-loop} approximation is motivated by KPZ-DDK and
 proved to be {\it exact} 
by using the {method of singular products} reviewed in Sect.~\ref{s:4} below.


\section{Algebraic check of DDK}

\begin{flushright}
\parbox[t]{5.4cm}
{Salieri:  \mbox{}\\
{\small ``I checked the harmony with algebra.
Then finally proficient in the science, 
I~risked the rare delights of creativity.''}
\\[.2cm]
{\sc A. Pushkin}, {\it Mozart and Salieri}}   
\end{flushright}

\tit{One-loop propagator}  

The one-loop correction to the 
propagator $\LA\vp(-p) \vp(p)\RA$   
is diven by the diagrams
$$
\includegraphics[width=10cm]{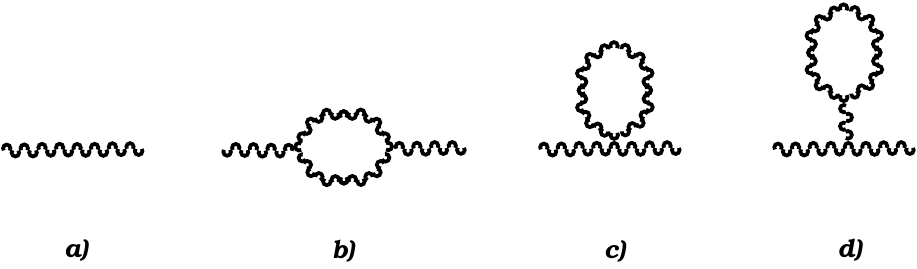} 
$$
Using the Pauli-Villars regularization with two regulators of mass squared  $M^2$ and one
regulator of mass squared $2M^2$ with normal statistics we find
\bea
\hbox{$b)$}&=&
-\frac14 \!\int \frac{\d^2 k}{(2\pi)^2} 
\left\{ \frac{\eps^2 k^2(k-p)^2}{(1+\eps k^2)[1+\eps(k-p)^2]}  
\right.\non && \hspace*{1.5cm}
-2 \frac{(\eps k^2 (k-p)^2-M^2)^2}{(k^2+M^2+\eps k^4)[(k-p)^2+M^2+\eps(k-p)^4]}
\non && \hspace*{1.5cm} \left.
+\frac{(\eps k^2 (k-p)^2-2M^2)^2}{(k^2+2M^2+\eps k^4)[(k-p)^2+2M^2+\eps(k-p)^4]}
\right.\} |\vp(p)|^2
\non
&\to&
\hbox{$b)$}\Big|_{\rm reg\ div}-\frac{p^2}{96\pi} |\vp(p)|^2 \nonumber
\eea

The one-loop {}{renormalization} of  $b^2$  where $T(\eps M^2)\sim \sqrt\eps M=$ tadpole $d) $
$$
\frac{1}{b^2}=\frac1{b_0^2}-
\left(\frac 16 -4 +T +2 {} G {}{\int \d k^2 \frac \eps{(1+\eps k^2)}}-
\frac 12 {} G T\right)+{\cal O}(b_0^2) 
$$
is ugly and involves a logarithmic divergence.

\tit{One-loop renormalization of $T_{zz}$}  

An analogous one-loop  {}{renormalization} of $T_{zz}^{(1)}$  is given by the diagrams
\vspace*{1mm}
$$
\hspace*{-.6cm}
\includegraphics[width=10.3cm]{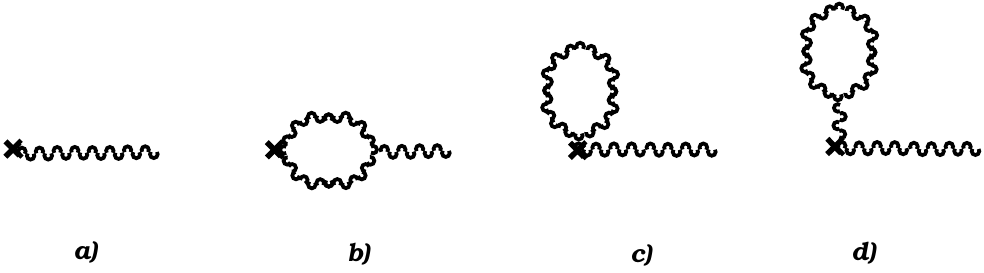} 
$$
leading again to the ugly result
$$
\frac q{b^2}=\frac{1}{b_0^2}-\frac 16 +2 -\frac 12 T -\frac 12{}  G 
- {} G {}{\int \d k^2 \frac {\eps }{(1+\eps k^2)} }+\frac 14 {} G T +{\cal O}(b_0^2)
$$

But squaring and multiplying by $b^2$, we find~\cite{Mak22}
$$
\frac {q^2}{b^2} =\left(\frac q{b^2}\right)^2\times b^2
=\frac{1}{b_0^2} -\frac 16 -{} G +{\cal O}(b_0^2)
$$
where the regularization-dependent parts as well as the logarithmic divergence goes away.
What remains precisely {}{confirms} the above shift of the central charge by ${}{6G}$
obtained by  the CFT technique. This is an example 
of tremendous cancellations  
that make Intelligent one  (ione) loop
 to  be exact leading to  
\bea
q -b^2=1 \label{DDK1} \\
-\frac6{b_0^2}  +\frac{6q^2}{b^2} +1+{}{6 G} q 
=0   \label{DDK2}
\eea
The solution to these equations is given in Sect.~\ref{s:6} below.

\section{Method of singular products
as pragmatic mixture\\[1mm]
of CFT and QFT\label{s:4}}
 
\tit{Generator of conformal transformation}

The conservation of Improved EMT is due to the classical equations of motion
$$
\bp T_{zz} =-\pi  \p \frac {\delta S}{\delta \vp} + \pi \p \vp\frac {\delta S}{\delta \vp}
$$
Thanks to the (quantum) Schwinger-Dyson equation
\be
\frac {\delta S}{\delta \vp} \stackrel{{\rm w.s.}}=\frac {\delta}{\delta \vp}
\qquad \hbox{or} \qquad
\LA\frac {\delta S}{\delta \vp} X\RA =\LA \frac {\delta X}{\delta \vp} \RA
\label{SD}
\ee
the generator of conformal transformation for a  nonquadratic EMT is~\cite{Mak22c}
\be
\hat \delta_\xi \equiv  \int_{C_1}\frac{\d z}{2\pi \i} \xi (z) T_{zz} (z) =
\frac1\pi\int_{D_1} \xi \bp T_{zz}
\stackrel{{\rm w.s.}}
=\int_{D_1} \left( \xi' \frac {\delta}{\delta \vp} +\xi \p \vp\frac {\delta }{\delta \vp}\right)
\label{hatdel}
\ee
where  $D_1$ includes  singularities of $\xi(z)$ and $C_1$ bounds $D_1$.  


Here the form of $\hat \delta_\xi$ in the middle is Noether's and
the standard form on the left  may involve
 many terms like for $T_{zz}$ given by \eq{Tab}.
The one on the right 
takes into account a tremendous cancellation of diagrams 
owing to the Schwinger-Dyson equation, but there are subtleties
 associated with singular products because the argument of the variational derivative is 
 inside $D_1$. The 1st term with $\xi'$ there is linked to the conformal weight  =1 of $\e^\vp$
  and is missing for the minimal EMT.  

 \tit{List of  simplest singular products}

The simplest singular product (which  {naively vanishes}) is
\be
\frac1{b^2}\int \d^2 z\, \xi(z) \LA \partial^n \vp(z) \vp(0) \RA 
\delta^{(2)}(z) =(-1)^{n}  \frac{2}{n(n+1)} \xi^{(n)}(0) 
\label{H0}
\ee
It arises already in a free CFT  by the formulas
$$
\delta^{(2)}(z) =\bp \frac {1}{\pi z},\qquad
\frac 1{z^n} \bp \frac 1z=(-1)^n \frac1{(n+1)!} \p^n \bp \frac 1z
$$

Alternatively it can be derived introducing  the regularization by $\eps$
$$
G_{\eps}(k)=\frac{1}{k^2(1+\eps k^2)},\qquad 
\delta^{(2)}_\eps(k)=\frac{1}{(1+\eps k^2)}
$$
We then have as $\eps\to0$
$$
8\pi
\int \d^2 z f(z) \p^{n} G_\eps(z)
\delta^{(2)}_\eps(z)= (-1)^n \frac 2{n(n+1)}
\p^n f(0)
$$
and additionally
$$
8\pi 
\int \d^2 z f(z)\big[ -4 \eps\p^{n+1} \bp G_\eps(z)\big]
\delta^{(2)}_\eps(z) = (-1)^n \frac2{(n+1)}
\p^n f(0)
$$

\tit{Method of singular products: some results}

The method of singular products allowed us to prove Intelligent one loop~\cite{Mak23c}.
The key points are simplicity of the calculation (Mathematica applied) and the fact that
$\vp$ enters $\hat \delta_\xi$ only linearly, so just one operator is renormalized providing
$T_{zz}$ is normal ordered. We obtained
$$
\hat \delta_\xi  \e^\vp\stackrel{\rm w.s.}=d^{(\e^\vp)}\xi' \e^\vp+ \xi \p \e^\vp\qquad
d^{(\e^\vp)}=q -b^2
$$
 for the exact conformal weight
and
$$
 \LA \hat \delta_\xi T_{zz} (\om)\RA = \frac{ c^{(\vp)}} {12}\xi'''(\om)
 \qquad    c^{(\vp)}=\frac{6q^2}{b^2}+1 +{}{6 G} q 
$$
 for exact central charge $c^{(\vp)}$ of $\vp$.
Equations \rf{DDK1}, \rf{DDK2} are thus exact.

One more result is 
(almost) the proof of {}{Universality} for the {Polyakov} string, i.e.\  of the fact that
the results do not depend on {}{higher-order} terms in the effective action
 {(shown for four- and six-derivative actions)}.

 \tit{Universality of higher orders for Polyakov string }

To show the {universality} at order $ {a^{2m}}$ ($2m\!+\!2$ -derivative terms) we use
$$
G_\eps(k)=\frac1{k^2(1+\eps k^2)^m}, \qquad 
\delta_\eps^{(2)}=\frac1{(1+\eps k^2)^m}
$$
with $m=1$ for the four-derivative action. The {singular product}
$$
8\pi \int \d^2 z \, \xi(z) \big[\p^n(-4\eps\p\bp)^k G_\eps(z) \big]\delta_\eps^{(2)}(z) =(-1)^n
H^{(k)}_{n,m}\p^n\xi(0)
$$
$$
H^{(k)}_{n,m}=2\frac{\Gamma(n+k)\Gamma(2m-k)\Gamma(m+n)}{\Gamma(n+1)\Gamma(m)\Gamma(2m+n)} 
$$ 
can be used for proving {}{universality} of $c$ and {}{marginality} 
of higher terms emerging for {}{Polyakov} string \ ({in support of the BPZ-assertion} mentioned
in Sect.~\ref{s:2}):
\bea
\LA \hat \delta_\xi T_{zz}^{(\vp,2)}(\om) \RA &=&
\frac {\xi'''(\om)}{2} \left(\frac{q^2}{b ^2}+ H^{(0)}_{2,m}-H^{(0)}_{3,m} \right) =
\xi'''(\om) \left(\frac{q^2}{2b ^2}+ \frac1{12} \right)\non
\LA \hat \delta_\xi T_{zz}^{(\vp,4)}(\om) \RA &=&
\frac  {\xi'''(\om) }4 \left( -2 H^{(1)}_{0,m} +2 H^{(1)}_{1,m} +6 H^{(1)}_{2,m}
 -4 H^{(1)}_{3,m}\right)=0 \non
\LA \hat \delta_\xi T_{zz}^{(\vp,6)}(\om) \RA &=&
\xi'''(\om)\left(-\frac 12 H^{(2)}_{0,m}- H^{(2)}_{1,m}+3 H^{(2)}_{2,m}
-\frac32  H^{(2)}_{3,m}\right)
=0 \nonumber
\eea
The same results hold for a more general  regularization.


\tit{Heuristic QFT understanding of universality}

For a {general action} 
\ \ \ \   ($F(x)=(1+x)/{8\pi b_0^2}$ for the four-derivative action)
$$
{S}^{\rm gen} [\vp]= -\frac 1{2}\int  \sqrt{\hg} 
 \vp \hat\Delta F( -\eps\e^{-\vp}\hat\Delta) \vp,\qquad F(0)=\frac 1{8\pi b_0^2}
$$
we have  $k^2 F(\eps k^2)$ for the {propagator   and $\eps k^4F'(\eps k^2)$ for
the triple vertex. 

The one-loop renormalization $\e^\vp \Rightarrow \e^{\alpha \vp}$ is given by
(wavy lines represent $\vp$)\\[5mm]
\centerline{\includegraphics[width=7cm]{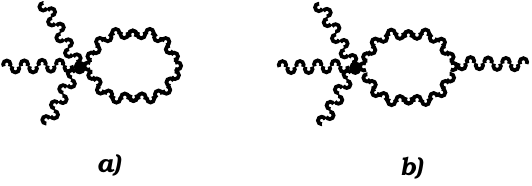}}
For the diagram b) we obtain
$$
b)=
\frac {\e^\vp}2
\times \vp \int \frac {\d^2 k}{(2\pi)^2} 
\frac {\eps k^4 F'(\eps k^2)}{[k^2F(\eps k^2)]^2}=  
\frac 1{8\pi F(0)}\e^\vp \vp =  \e^\vp b^2_0 \vp 
$$
{}{independently} on  the choice of $F$ like 
anomalies in QFT.

\section{Quantum conformal symmetry}

\tit{QFT Virasoro algebra?}

The commutator of two conformal transformations is~\cite{Mak22c} \hfill($\zeta= \xi\eta'-\xi'\eta$)
\bea
&&\LA(\hat \delta_\eta \hat \delta_\xi -\hat \delta_\xi \hat \delta_\eta) X \RA =
\LA \hat \delta_\zeta  X \RA + \int_{D_1}\!\d^2 z \int_{D_z} \!\d^2 \om \non &&\hspace*{.5cm}
\times\LA\big[ q\xi'(z)+\xi(z) \p \vp(z) \big] \big[ q\eta'(\om)+\eta(\om) \p \vp(\om) \big]
\frac{\delta^{2} S}{\delta\vp(z) \delta \vp(\om)} X\RA~ \label{Hess}
\eea
where  $D_1$  includes singularities of $\xi(z)$ and $\eta(z)$, 
and $D_z$  comprises $z$ = the commutator in  operator formalism.
The {Schwinger-Dyson} equation~\rf{SD} is used in the derivation of \eq{Hess}.

The 1st term on the right-hand side of \eq{Hess} is pure classical, 
while the 2nd term which can always be reduced 
to the standard form
$$
\LA(\hat \delta_\eta \hat \delta_\xi -\hat \delta_\xi \hat \delta_\eta) X \RA=
\LA \hat \delta_\zeta  X \RA+ \frac{1}{24} \oint _{C _1}\frac{\d z}{2\pi \i} \big[ \xi'''(z)\eta(z)-\xi(z)\eta'''(z) \big]
\Big\langle {}{\hat c}X \Big\rangle
$$
is  linked to a central charge of the {}{Virasoro} algebra. 
${}{\hat c}$ is the usual c-number for quadratic actions, reproducing the usual case,
but ${}{\hat c}$ may be $\vp$-dependent for the four-derivative actions  with $G\neq 0$,
which would generalize the Virasoro algebra to QFT.

As we shall now demonstrate,  
in Intelligent one-loop approximation ${\hat c}$ is c-numbered and the Virasoro
algebra is as usual.

 \tit{Central charge at one loop}
 
 To deal with the four-derivative action we expand in  powers of $\vp$ 
$$
{\cal S}={\cal S}^{(2)}+{\cal S}^{(3)}+{\cal S}^{(4)}+{\cal O}(\vp^5)
$$
The Hessian for the quadratic part of the action reads
$$
\frac{\delta^{2} S^{(2)}}{\delta\vp(z) \delta \vp(\om)}=\frac{1}{8\pi b^2} (-4\p\bp) \delta^{(2)}(z-\om)
$$
yielding
\bea
 \int_{D_1}\! \!\d^2 z \int_{D_z} \!\!\d^2 \om \,
q^2\xi'(z)\eta'(\om)
\frac{\delta^{2} S^{(2)}}{\delta\vp(z) \delta \vp(\om)}
=\frac{q^2}{4b^2} \oint _{C _1}\frac{\d z}{2\pi \i} \big[ \xi'''(z)\eta(z)-\xi(z)\eta'''(z) \big]
\nonumber
\eea
and
\bea
 \lefteqn{\int_{D_1}\! \!\d^2 z \int_{D_z} \!\!\d^2 \om \,\xi(z)\eta(\om)
\LA \p \vp(z)  \p \vp(\om) \RA
\frac{\delta^{2} S^{(2)}}{\delta\vp(z) \delta \vp(\om)} }\non
&&=-\frac 1{2b^2} \oint _{C _1}\frac{\d z}{2\pi \i} \xi(z) \int _{D_z}\d^2\om
 \big[ \eta'(z)\LA \p^2\vp(\om) \vp(z) \RA
+\eta(z)\LA \p^3\vp(\om) \vp(z) \RA\big] 
\non
&&~~\times\delta^{(2)}(z-\om)
=\frac{1}{24} \oint _{C _1}\frac{\d z}{2\pi \i} \big[ \xi'''(\om)\eta(z)-\xi(\om)\eta'''(z) \big]  
\nonumber
\eea
where we have used the formula~\rf{H0} for the singular products. 
The sum of two gives  $c=6 q^2 /b^2+1$, reproducing    DDK.


The average in the 2nd term in \eq{Hess} factorizes at one loop and we obtain 
\bea
 \lefteqn{\int_{D_1}\!\d^2 z \int_{D_z} \!\d^2 \om \,
\LA\big[\xi'(z)\eta(\om) \p\vp(\om)+ \xi (z)\eta'(\om) \p\vp(z)\big]
\frac{\delta^{2} S^{(3)}}{\delta\vp(z) \delta \vp(\om)} \RA}\non
&&=\frac{G\eps}{b^2}  \oint _{C _1}\frac{\d z}{2\pi \i} 
\int_{D_z} \d^2 \om \big[ \xi''(z)\eta(\om)+\xi(z)\eta''(\om) \big] 
\LA \p^2\bp \vp(\om)\vp(z) \RA \delta^{(2)} (\om-z)
\non 
&&=\frac{G}{4} \oint _{C _1}\frac{\d z}{2\pi \i} \big[ \xi'''(z)\eta(z)-\xi(z)\eta'''(z) \big]
\nonumber
\eea
It remarkably reproduces $\delta {c}=6G$ at one loop \ ${}{\Longrightarrow}$  
\ the 4th check!

This is not yet the whole story because of
\bea
 &&\int_{D_1}\!\d^2 z \int_{D_z} \!\d^2 \om \,
\xi'(z)\eta'(\om) \LA
\frac{\delta^{2} S^{(4)}}{\delta\vp(z) \delta \vp(\om)} \RA
\non 
&&\hspace*{1cm}=\frac{G}{4} \oint _{C _1}\frac{\d z}{2\pi \i} \big[ \xi'''(z)\eta(z)-\xi(z)\eta'''(z) \big]
\left(  -2  \int \frac{\eps \d k^2}{1+\eps k^2}  \right)
\nonumber
\eea
which logarithmically diverges, so the contribution of $S^{(3)}+S^{(4)}$ to $\delta c$
 is positive for negative $G$ in accordance with $c$-theorem.
However, the divergence is mutually canceled with the
one arising in the calculation of observables, like in that of the string susceptibility.
These logarithmic divergences never appear in the method of singular products if
$T_{zz}$ and ${S}$ are normal ordered.

\section{Relation to minimal models \label{s:6}} 
 
\tit{Exact solution for four-derivative action}

{The solution} to  two modified {DDK} equations~\rf{DDK1}, \rf{DDK2}   
with $b_0^2=6/(26-d)$ is
\bea
b^{-1}&=&\sqrt{\frac{d_+-d}{24}}+\sqrt{\frac{d_--d}{24}}\label{b}\\
 d_\pm&=&13-6 G\pm 12 \sqrt{1+ G} 
\label{dpm}
\eea

The {KPZ} barriers were $d_-=1$, $d_+=25$ for the Polyakov string 
associated with $G=0$ but
now they are shifted to $d_\pm$ given by \eq{dpm} which depend on $G\in [-1,0]$. 
For the Nambu-Goto string ${G=-1/3}$  ${}{\Longrightarrow}$ 
$
d_-=15-4\sqrt{6}\approx 5.2 >4
$.

The {}{string susceptibility} equals
$$
\gamma_{\rm str}=(h-1) \frac {q}{b^2}+2 =
(h-1)\frac{25-d-6 G+\sqrt{(d_+-d)(d_--d)}}{12}+2 
$$
It is {}{real} for $d<d_-$ with $d_->1$ increasing from 1 at $G=0$ to 19 at $G=-1$
(the range $0\geq G\geq-1$  required for {}{stability}
{}{as it follows   from the identity~\rf{modulo}).


\tit{BPZ null-vectors and Kac's spectrum}

Like in the usual {}{Liouville} theory the operators 
$$
V_\al=\e^{\al \varphi}, \qquad \al=\frac{1-n}2+\frac{1-m}{2b^2}
$$
{\'a la} Dotsenko-Fateev~\cite{DF84}
are the {}{BPZ null-vectors} for integer $n$ and $m$,  obeying 
$$
\left( L^2_{-1} +b^2 L_{-2} \right) \e^{-\vp/2} \stackrel{{\rm w.s.}}=0, 
\quad
\left( L^2_{-1} +b^{-2} L_{-2} \right) \e^{-b^{-2}\vp/2} \stackrel{{\rm w.s.}}=0, \quad \ldots
$$
 as $\eps\to0$. Their conformal weights 
$$
\Delta_\al =\al+(\al-\al^2) b^2
$$
coincide~\cite{Mak23e} with  Kac's spectrum  
 $$
\Delta_{m,n}({} c)=\frac{ c-1}{24} +\frac 14 \left( (m+n)\sqrt{\frac {1- c}{24}}+
 (m-n)\sqrt{\frac {25- c}{24} } \right)^2
$$
providing the central charge of  the Virasoro algebra equals
$$
{} c=26-d+G\frac{\left[25-d-6  G+\sqrt{(d_+-d)(d_--d)}\right]}{2(1+G)} =1+6(b+b^{-1})^2
$$
reproducing the standard relation between $c$ and $b$ given by \eq{b}.

\tit{Minimal models from four-derivative action}

To describe {}{minimal models} we choose like in the usual {}{Liouville} theory
\be
c=25+6 \frac {(p-q)^2}{pq} \quad {}\Longrightarrow\quad
G=\frac {(1-d-6 \frac {(p-q)^2}{pq})q}{6(q+p)}
\label{G}
\ee
with coprime $q>p$. 
If $G=0$ this would imply  (the {KPZ} result)
$$d
=1-6 \frac {(p-q)^2}{pq} \quad{}{\Longleftarrow}\quad G=0$$
 for the central charge of {matter} but now $d$ is 
a {free} parameter obeying
$$
1-6\frac{(p-q)^2}{pq}\leq d\leq 19-6\frac pq \quad {}\Longleftarrow
\quad
0\geq G \geq -1
$$
Contrary to the usual {}{Liouville} theory now {}{Kac's} $ {} c\neq c^{(\vp)}=26-d$.

 Remarkably, \eq{G} is satisfied for
 ${G=-1/3}$  in $d=4$ by  $p=3$, $q=p+1=4$ which 
 is associated~\cite{Mak24}  with the (4,3) unitary minimal model
 like the critical {Ising} model on a random lattice.
 
 My final comment is about the critical indices which are determined by the value of $b^2$.
 For the perturbative branch $b^2=p/q$ like in the usual
Liouville theory but domain of applicability is now {}{broader} encouraging 
numerous applications in Statistical Mechanics \'a la
Kogan-Mudry-Tsvelik~\cite{KMT96}.

\section{From Nambu-Goto string to four-derivative action \label{s:8}} 

\tit{Nambu-Goto vs.\ Polyakov strings}

The action of the {Nambu-Goto string} is the area of the string worldsheet.
By introducing the 
(imaginary) {}{Lagrange multiplier} $\lambda^{ab}$ and an {independent}
metric tensor $g_{ab}$,  it can be rewritten in the form
\be
K_0 \!\int \d^2\omega\,\sqrt{\det \partial_a X \cdot \partial_bX}=
K_0 \!\int \d^2\omega\,\sqrt{ g}  
+\frac{K_0}2 \!\int \d^2\omega\, \lambda^{ab} \left( \partial_a X \cdot \partial_bX -g_{ab}
\right)
\label{NG}
\ee
It is convenient to use the worldsheet coordinates  
$\omega_1,\omega_2 \in  $  
rectangle. 

The {ground state} of the 
action~\rf{NG} is at $\lambda^{ab}=\blambda \sqrt{g}g^{ab}$.
Classically $\blambda_{\rm cl}=1$ 
${} \Longrightarrow$ {Polyakov's string} 
$$
{\cal S}=
\frac{K_0}2 \!\int \d^2\omega\, \sqrt{g}g^{ab}  \partial_a X \cdot \partial_bX
$$
We consider the kinematics of a closed bosonic string winding once around {}{compactified}  dimension
of circumference $\beta$ and propagating (Euclidean) time $L$. 
The topology is of cylinder or torus (bagel).
There is no tachyon} if $\beta$ is {}{large enough}.

The Gaussian path integral 
over $X^\mu$ with the action~\rf{NG} can be done by splitting
$X^\mu=X^\mu_{\rm cl} +X^\mu_{\rm q}$.
This produces an {emergent} (or {}{effective}) action
\bea
S[g_{ab},\lambda^{ab}]&=& K_0 \int \d^2\omega\,\sqrt{g} 
+\frac{K_0}2 \int \d^2\omega\, \lambda^{ab} 
\left( \partial_a X_{\rm cl } \cdot \partial_bX_{\rm cl } -g_{ab} \right)  \non
&&+ \frac{d}{2}  \tr \log {\left( -\frac1 {\sqrt{g} }  \partial _a \lambda^{ab} \partial_b\right)}
+~\hbox{ghost contribution}
\nonumber
\eea
where the contribution from the ghosts is the usual one.

\tit{Generalized conformal anomaly}

Path-integrating out $X^\mu_q$ (and the ghosts), we arrive at~\cite{Mak23c}
\bea
{S}[g_{ab}, \lambda^{ab}]= K_0 \int \left(\sqrt{g}
-\frac12  \lambda^{ab}g_{ab} \right)
+ \frac 1{8\pi \ta} \int \sqrt{g}\left[ -
\frac {d}{\sqrt{\det{\lambda^{ab}}}}+2\right]
\non
+ \frac 1{96\pi} \int \left[ (d-26)\sqrt{g}\,R\frac1 \Delta R - d \left(
\beta\lambda^{ab} g_{ab} R+2
\lambda^{ab}\nabla_a\p_b \frac1\Delta R  \right)
\right]
\label{S}
\eea
where higher orders in the ultraviolet cutoff $\ta$ are dropped.
For the Nambu-Goto string $\beta=1$ but we keep it {arbitrary} for generality.

The emergent action~\rf{S} is derived  from the {}{DeWitt-Seeley} expansion of 
$$
{\cal O}\equiv-\frac 1 {\sqrt{g}}\p_a \lambda^{ab} \p_b = -h^{ab} \p_a\p_b + A^a \p_a,\quad
h^{ab}=\frac{\e^{-\vp}}{\sqrt{\hat g}} \lambda^{ab}
,~~ A^b=\frac {\e^{-\vp} }{\sqrt{\hat g}} \p_a \lambda^{ab}
$$
\vspace*{-5mm}
$$
\LA \left| \e^{-\tau {\cal O} }
\right|  \RA = \frac 1{4\pi\tau}
 +\frac1{4\pi} \left(\frac 16{R(h)} +E(A,h)
 \right)+{O}(\tau)$$
 
The same  result can be derived, calculating   the {Coleman-Weinberg} effective action
with the use of the {Pauli-Villars} regularization in the diagrams
\be
\frac {d}2\,{\rm tr}\ln \left[-\frac1{{\rho}}\partial _a \lambda^{ab}\partial_b
\right] _{\rm reg}~=~
\sum_n \frac 1n \!\!\!\!
\unitlength=.5mm
\linethickness{1.6pt}
\begin{picture}(37.00,14.50)(1,18.6)
\put(20.00,20.00){\circle{10.00}}
\multiput(16.46,23.535)(-2.00,3.33){3}{
  \bezier{28}(0.00,0.00)(-1.333,0.333)(-1.00,1.665)}
\multiput(15.46,25.20)(-2.00,3.33){2}{
  \bezier{28}(0.00,0.00)(0.333,1.167)(-1.00,1.665)}
\multiput(23.535,23.535)(2.00,3.33){3}{
  \bezier{28}(0.00,0.00)(1.333,0.333)(1.00,1.665)}
\multiput(24.535,25.20)(2.00,3.33){2}{
  \bezier{28}(0.00,0.00)(0.333,1.167)(1.00,1.665)}
\put(20.00,30.00){\makebox(0,0)[cc]{$\cdot $}}
\put(23.00,29.50){\makebox(0,0)[cc]{$\cdot $}}
\put(17.00,29.50){\makebox(0,0)[cc]{$\cdot $}}
\put(11.00,15.50){\makebox(0,0)[cc]{$\cdot $}}
\put(9.80,18.50){\makebox(0,0)[cc]{$\cdot $}}
\put(12.90,13.00){\makebox(0,0)[cc]{$\cdot $}}
\put(27.10,13.00){\makebox(0,0)[cc]{$\cdot $}}
\put(29.00,15.50){\makebox(0,0)[cc]{$\cdot $}}
\put(30.20,18.50){\makebox(0,0)[cc]{$\cdot $}}
\multiput(20.00,15.00)(0.00,-4.00){2}{
  \bezier{28}(0.00,0.00)(-1.00,-1.00)(0.00,-2.00)}
\multiput(20.00,13.00)(0.00,-4.00){2}{
  \bezier{28}(0.00,0.00)(1.00,-1.00)(0.00,-2.00)}
\end{picture}
\ee
where the wavy lines correspond to fluctuations $\delta \lambda^{ab}$
or $\delta g_{ab}$ about the ground state.

\tit{Conformal gauge and flat background}

The emergent action~\rf{S} becomes local in the {conformal gauge}~\rf{confog}
Choosing  {conformal coordinates} $z$ and $\bz$ in  a flat background 
$g_{zz}=g_{\bz\bz}=0$, $g_{z\bz}=g_{\bz z}=1/2$, we arrive at Euclidean CFT
{(remember $\beta=1$ for the Nambu-Goto string)}
\bea
{\cal S}[\vp, \lambda^{ab}]
&=&K_0\int \e^\vp(1-\lambda^{z\bz})
+\frac {1}{24\pi} \int \Big\{ \frac {3\e^\vp}{\ta} 
\Big(2- \frac {d}{\sqrt{\det{\lambda^{ab}}}} \Big)\non && \hspace{-1cm}
+(d-26)\vp\p\bp\vp 
+d \big[2(1+\beta)\lambda^{z\bz} \p\bp \vp +
\lambda^{zz}\nabla\p \vp +\lambda^{\bz\bz}\bar\nabla\bp\vp \big]
\Big\} \label{Se}
\eea
where $\nabla = \p-\p \vp$ is the {covariant derivative} in  conformal gauge, so it
 describes the theory with an interaction. 
 (There is no such interaction for large $\beta$ when 
 only $\lambda^{z\bz}=\lambda^{ab}\hg_{ab}$ remains, reproducing the case~\cite{KN93}.)
 
 Expanding  the square root in \rf{Se} in the fluctuations $\delta\lambda^{ab}$,
 we see they have mass squared  $\sim a^{-2}$ and are thus  
{localized}~\cite{Pol87}.
Therefore, {only $\vp$ fluctuates} (the fluctuations are stable for $d<26$) 
${\Longrightarrow}$ the Liouville theory.
However, the private life of $\lambda^{ab}$s which occurs 
at distances $\sim  {a}$ 
is nevertheless macroscopically observable as demonstrated above.

 \tit{Equivalence with four-derivative Liouville action}

The path integral over $\delta\lambda^{ab}$ has  a {}{saddle point} justified 
by small $\ta$ at
\be
\delta\lambda^{ab} = \frac{1}{3}\sqrt{g}\ta \left(g^{ac}g^{bd}\nabla_c\p_d \vp
+\frac{(\beta-1)}4 g^{ab} \Delta\vp\right) +{\cal O}(\taa)
\label{lll}
\ee
We thus  arrive at  the four-derivative Liouville action~\rf{4dLiou} modulo the terms
${\cal O}(\ta)$ which are omitted by construction. 
The parameters of the action are
$$
  b^2_0=\frac 6{26-d},\quad G=-\frac1{1+(1+\beta)^2/2}, \quad
\eps=\frac {2d\blambda^3 } {3G(d-26)}\ta \sim \Lambda^{-2}
$$
resulting in  {$G=-1/3$}  for the {}{Nambu-Goto} string where {$\beta=1$}.

 \tit{Emergence of nonlocal term in flat space}

The nonlocal term in \eq{Tab} can be obtained directly in flat space from
the energy-momentum tensor of the Nambu-Goto string
$$
T^{({\rm NG})}_{zz}=2\pi \big(\lambda^{z\bz}\p X\cdot \p X +\lambda^{\bz\bz}\p X\cdot \bp X  \big)+
2\pi \big(\lambda^{z\bz}\p Y\cdot \p Y +\lambda^{\bz\bz}\p Y\cdot \bp Y  \big)
$$
Averaging over $X^\mu$ and the regulators $Y^\mu$, we find

$$
\hspace*{-7.cm}T_{zz} =\LA T^{({\rm NG})}_{zz} \RA_{X,Y}=
$$
$$
\includegraphics[width=9.6cm]{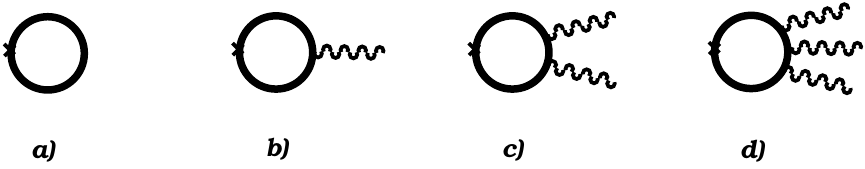}
$$
where the solid line represents $X^\mu$ (or $Y^\mu$) 
and the wavy lines represent $\delta\lambda^{ab}$ or $\vp$.

Thanks to the interaction $ \lambda^{zz}\p X\cdot \p X$ the
dagram $b)$ produces  $p_z^4/p^2 \lambda^{zz}$ (because $X^\mu$ is massless):
\bea
b)&=&d\int \d^2 \om\big[\p^2 G_0(\om-\om_0)\big]^2 \lambda^{zz}(\om) =
d\int \d^2 \om\frac1{(\om-\om_0)^4}\lambda^{zz}(\om) \non
&=&d\,\p^3 \int \d^2 \om\frac1{(\om-\om_0)}\lambda^{zz}(\om) \equiv
d \,\p^3 \frac1{\bp}\lambda^{zz}(\om_0) \nonumber
\eea
Using \eq{lll}, this reproduces the nonlocal term in the $T_{zz}$ component of \rf{Tab}.

Likewise, the
dagram $b)$ produces $p_z^2\vp$  for the regulator $Y^\mu$  thanks to 
the interaction $M^2 \vp Y^2$
in \rf{Sreg}. This is precisely the linear part of Improved EMT in the Liouville theory.

\section*{Conclusion}

\begin{itemize}
\item``Massive'' CFTs exist and  are solvable by the (almost) usual CFT technique 

\item MCFTs allow higher derivatives and are intimately linked to Improved EMT
involving generically {a} {nonlocal} term 

\item {It is crucial} that Improved  EMT is conserved and traceless 
(which is specific to theories with diffeomorphism invariance)

\item {The nonlocal} term has a long tail $\Rightarrow$ magnifying small distances and 
making the four-derivative term observable

\item {The Nambu-Goto} and {Polyakov} strings are {told apart} by higher-derivative 
terms which revive quantumly like anomalies in QFT

\item The violation of {}the {BPZ-assertion} for the {}{Nambu-Goto} string is due to nonlocality 
({the Virasoro algebra is modified classically by a  nonlocal contribution})

\item The emergence of the four-derivative {}{Liouville} action 
from the Nambu-Goto string alludes to a (4,3) 
\mbox{minimal} model
like  for the critical {}{Ising} model on a random lattice 

\end{itemize}

{My warmest  CONGRATULATIONS  to Irina on her Jubileum!}

\vfill


\begin{thebibliography}{10}

\bibitem{KPZ88} 
  V.G.~Knizhnik, A.M.~Polyakov and A.B.~Zamolodchikov,
  {\it Fractal structure of 2D quantum gravity,}
  Mod.\ Phys.\ Lett.\ A {\bf 3} (1988)  819.
\bibitem{Dav88} 
F.~David,
{\it Conformal field theories coupled to 2D gravity in the conformal gauge,}
  Mod.\ Phys.\ Lett.\ A {\bf 3} (1988)  1651.
\bibitem{DK89} 
  J.~Distler and H.~Kawai,
  {\it Conformal field theory and 2D quantum gravity,} 
  Nucl.\ Phys.\ B {\bf 321} (1989) 509.
\bibitem{AM16b}
J.~Ambjorn and Y.~Makeenko,
{\it String theory as a Lilliputian world},
Phys.\ Lett.\ B {\bf 756}  (2016) 142
[arXiv:1601.00540  [hep-th]].

\bibitem{Mak22}
Y.~Makeenko,
{\it Opus on conformal symmetry of the Nambu-Goto versus Polyakov strings},  
Int. J. Mod.\ Phys.\ A {\bf 38} (2023) 2350010
[arXiv:2204.10205   [hep-th]].

\bibitem{CCJ70}
C.G. Callan, S.R. Coleman and R. Jackiw,
{\it A new improved energy-momentum tensor},
Ann.\ Phys. {\bf 59} (1970) 42.
  
\bibitem{DJ95}
 S. Deser and  R. Jackiw,
{\it Energy-momentum tensor improvements in two dimensions},
Int.\ J.\ Mod.\ Phys.\ B {\bf 10} (1996) 1499
[arXiv:9510145 [hep-th]].
  
\bibitem{KN93}
H. Kawai and R. Nakayama
{\it Quantum R**2 gravity in two-dimensions},
Phys.\ Lett.\ B {\bf 306} (1993) 224
[arXiv:hep-th/9303006 [hep-th]].
\bibitem{BPZ}
A.~Belavin, A.~Polyakov and A.~Zamolodchikov,
{\it Infinite conformal symmetry in two-dimensional quantum field theory},
Nucl.\ Phys.\ B {\bf 241}, 333 (1984).

\bibitem{Mak22c}
Y. Makeenko,
{\it Notes on higher-derivative conformal theory with nonprimary energy-momentum tensor
 that applies to the Nambu-Goto  string}, JHEP {\bf 01} (2023) 110 [arXiv:2212.02241 [hep-th]].

\bibitem{Mak23c} 
Y.~Makeenko,
{\it Singular products and universality in higher-derivative conformal theory},
JHEP {\bf 09} (2023)  086 [arXiv:2307.06295 [hep-th]].

\bibitem{DF84}
 V.S. Dotsenko and  V.A. Fateev,
{\it Conformal algebra and multipoint correlation functions in two-dimensional statistical models},
Nucl.\ Phys.\ B {\bf 240},  312 (1984).

\bibitem{Mak23e}
Y.~Makeenko,
{\it Exact solution of higher-derivative conformal theory and minimal models},
Phys.\ Lett.\  {\bf 845} (2023) 138170  [arXiv:2308.05030 [hep-th]].

\bibitem{Mak24} 
Y.~Makeenko,
{\it Nambu-Goto string as a higher-derivative Liouville theory},
Phys.\ Rev.\ D {\bf 110} (2024)  L081903
 [arXiv:2407.01136 [hep-th]].
 
 \bibitem{KMT96}
I. Kogan, C. Mudry and A. Tsvelik,
{\it The Liouville theory as a model for prelocalized states in disordered conductors},
Phys.\ Rev.\ Lett.\ {\bf 77} (1996) 707  
[arXiv:9602163 [cond-mat]].
\bibitem{Pol87}
A.M.~Polyakov, {\it Gauge fields and strings},  
Harwood Acad.\ Pub. (1987).

\end{thebibliography}
\end{document}